\documentclass[epj]{webofc}
\usepackage[varg]{txfonts}   
\usepackage{hyperref}

\renewcommand*{\eqref}[1]{Eq.~(\ref{eq:#1})}

\wocname{\includegraphics[width=0.25cm,clip]{logoARENA} ARENA2018}
\wocname{ARENA2018}
\woctitle{ARENA2018}
\woctitle{\includegraphics[width=0.25cm,clip]{logoARENA} ARENA2018}
\begin{document}
\title{Hydrophone characterization for the KM3NeT experiment}

\author{\firstname{Rasa} \lastname{Muller}\inst{1,3}\fnsep\thanks{\email{rasam@nikhef.nl}}
\and \firstname{Sander} \lastname{von Benda-Beckmann}\inst{2}
\and \firstname{Ed} \lastname{Doppenberg}\inst{1}
\and \firstname{Robert} \lastname{Lahmann}\inst{4}
\and \firstname{\mbox{Ernst-Jan}} \lastname{Buis}\inst{1}
\firstname{{on behalf of the KM3NeT collaboration}}}

\institute{TNO, Technical Sciences, Optics department, Delft, the Netherlands
           \and TNO, Technical Sciences, Acoustics and Sonar department, The Hague, the Netherlands
           \and University of Amsterdam, Amsterdam, the Netherlands
           \and Erlangen Center for Astroparticle Physics, University of Erlangen, Erlangen, Germany }

\abstract{With the KM3NeT experiment, which is presently under construction in the Mediterranean Sea, a new neutrino telescope will be installed to study both the neutrino properties as well as the cosmic origin of these particles. To do so, about 6000 optical modules will be installed in the abyss of the Mediterranean Sea to observe the Cherenkov radiation induced by high energy particle interactions in the deep sea.
As each module of the KM3NeT telescope includes a piezo hydrophone, KM3NeT will also provide a unique matrix of underwater hydrophones. Results from the measurements show a well understood response of continuous signals, such as tones. In contrast, the response to transients signals exhibit a complex behavior with ringing and echo's. Amplitude calibration measurements show a frequency dependent response which can be corrected for. Finally a system noise floor has been determined which amounts to 45 dB Re $\mu$Pa$^2$/Hz at 30 kHz.}

\maketitle
\section{Introduction}
\label{intro}
KM3NeT is designed to detect high-energy neutrino sources in the Universe as well as to determine the properties of neutrinos \cite{Loi, Elewyck2018}. Cherenkov radiation originating from neutrino induced particle showers is sensed using photo-multiplier tubes (PMTs). These PMTs are housed in so-called digital optical modules (DOM), of which 18 are attached to long strings, called detection units. Each detection unit (DU) will be anchored to the sea floor, and held vertical by a submerged buoy. Next to the PMTs, the payload of the optical modules includes various other sensors, such as accelerometers and a compass (to monitor the orientation of the module), as well as a hydrophone (to determine the position of the DOM using acoustic triangulation). 

By deploying a large number of detection units, a large submerged 3D grid will be established with a total detection volume of about 1 km$^3$, equiped with around 6000 optical modules. KM3NeT will therefore be one of the largest neutrino telescopes in terms of detection volume and number of sensors. Because each module includes a hydrophone it is interesting to assess the potential of KM3NeT in the acoustic detection of neutrinos \cite{Lah18}. To this end the piezo hydrophone in the DOM has been characterized and calibrated.

This paper is organized as follows. After a description of KM3NeT in the next section, the experimental setup and procedure for characterizing the piezo electric hydrophone will be discussed in section \ref{sec:experimental}. The results are presented in section \ref{sec:results}.
Finally, the conclusions and outlook for further work is discussed in section \ref{sec:conclusions}.

\section{The KM3NeT experiment}
\label{sec:KM3NeT}

Thanks to the large amount of PMTs per DOM, the neutrino induced shower can be reconstructed with a good angular resolution, but it requires accurate knowledge of the position of the DOMs. These positions need to be monitored on regular basis as marine currents cause relative dynamics in the telescope geometry. For this purpose an Acoustic Positioning System (APS) is developed, which is described in detail in \cite{Lar13}. The APS is based on trilateration and is expected to have a 10 cm accuracy \cite{Lar13, Vio16}. Besides piezo hydrophone receivers, dedicated transmitters beacons are designed for this purpose and will be deployed to the seafloor. The beacon's transmitting voltage response (TVR) is optimized in the range between 20 and 40 kHz.

\begin{figure}
  \centering
  \begin{minipage}[]{0.45\textwidth}
    \begin{center}
    \includegraphics[width=0.8\textwidth]{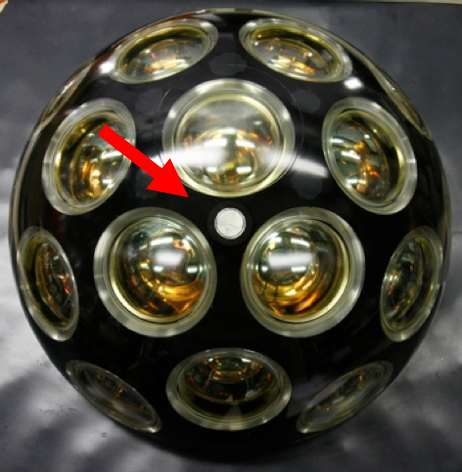}
      \end{center}
  \end{minipage}
  \begin{minipage}[]{0.45\textwidth}
    \begin{center}
    \includegraphics[width=0.8\textwidth]{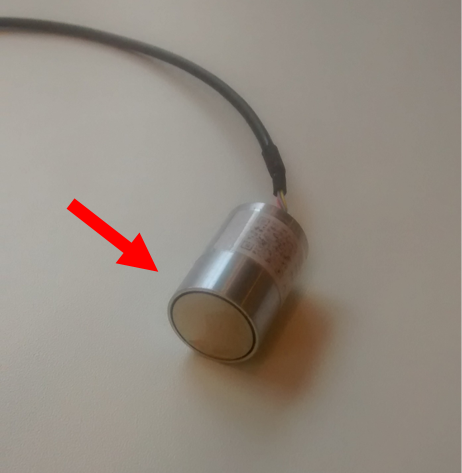}
      \end{center}
  \end{minipage}
  \vspace{0.5cm}\\
  \begin{minipage}[]{0.45\textwidth}
    \begin{center}
    (a)
      \end{center}
  \end{minipage}
  \begin{minipage}[]{0.45\textwidth}
    \begin{center}
    (b)
      \end{center}
    \end{minipage}
  \caption[]{(a) Photo of the digital optical module (DOM). The photo-multiplier tubes are clearly visible. The piezo ceramic hydrophone is visible as the small white dot. In (b) a picture of a hydrophone that is not (yet) integrated in the DOM. The piezo ceramic is indicated in both pictures.}
\label{fig:KM3NeT_piezo}
\end{figure}

The KM3NeT piezo hydrophone consists of a cylindrical shaped piezo ceramic sensor of 18 mm in diameter and 12 mm in height. A hydrophone is implemented in each DOM as shown in figure \ref{fig:KM3NeT_piezo}. A piezo ceramic crystal is connected to a pre-amplifier and integrated inside an aluminum tube (diameter 21 mm, height 30 mm) (figure \ref{fig:KM3NeT_piezo}). The aluminum tube serves as common ground for the acoustic sensor and ensures sufficient heat conduction to the glass sphere and thus to the deep sea environment \cite{Enz14b}. The sensor is glued directly onto the glass sphere. Signals from the piezo ceramic are transferred to an amplifier which houses in the aluminum cylinder close to the piezo. The signal is amplified, digitized using a 24 bit ADC and subsequently sent to the central electronics board inside the DOM where it is prepared to be sent through an optical fiber communication line to shore. The hydrophone has been designed and developed at the Erlangen Centre for Astroparticle Physics (ECAP), Germany.

\section{Experimental set-up}
\label{sec:experimental}
The experiment is performed with a single DOM in an anechoic basin at TNO, The Hague, the Netherlands. The 8m x 10m x 8m acoustically insulated basin is primarily designed for under water acoustic measurements and has a reflection loss of 6 - 16 dB. The DOM under study (production number 128) was assembled at Nikhef, Amsterdam, the Netherlands. As the DOM under test is not part of a complete assembled DU, a dedicated watertight tube was constructed to guide the optical fiber and copper cable (for communication and power respectively) to the data-acquisition system that was running on a laptop.

Sound signals were programmed, generated and emitted using an omni-directional transmitter of type ITC 1042. To characterize the response of the DOM hydrophone, the transmitted signal was measured using a reference hydrophone of type B$\&$K8101 as well. The reference signal is conditioned and sampled by a DAQ device (NXI, National Instruments) and subsequently sent to a laptop to analyze the result.

\begin{figure}[h]
  \centering
      \includegraphics[width=12cm, clip=true]{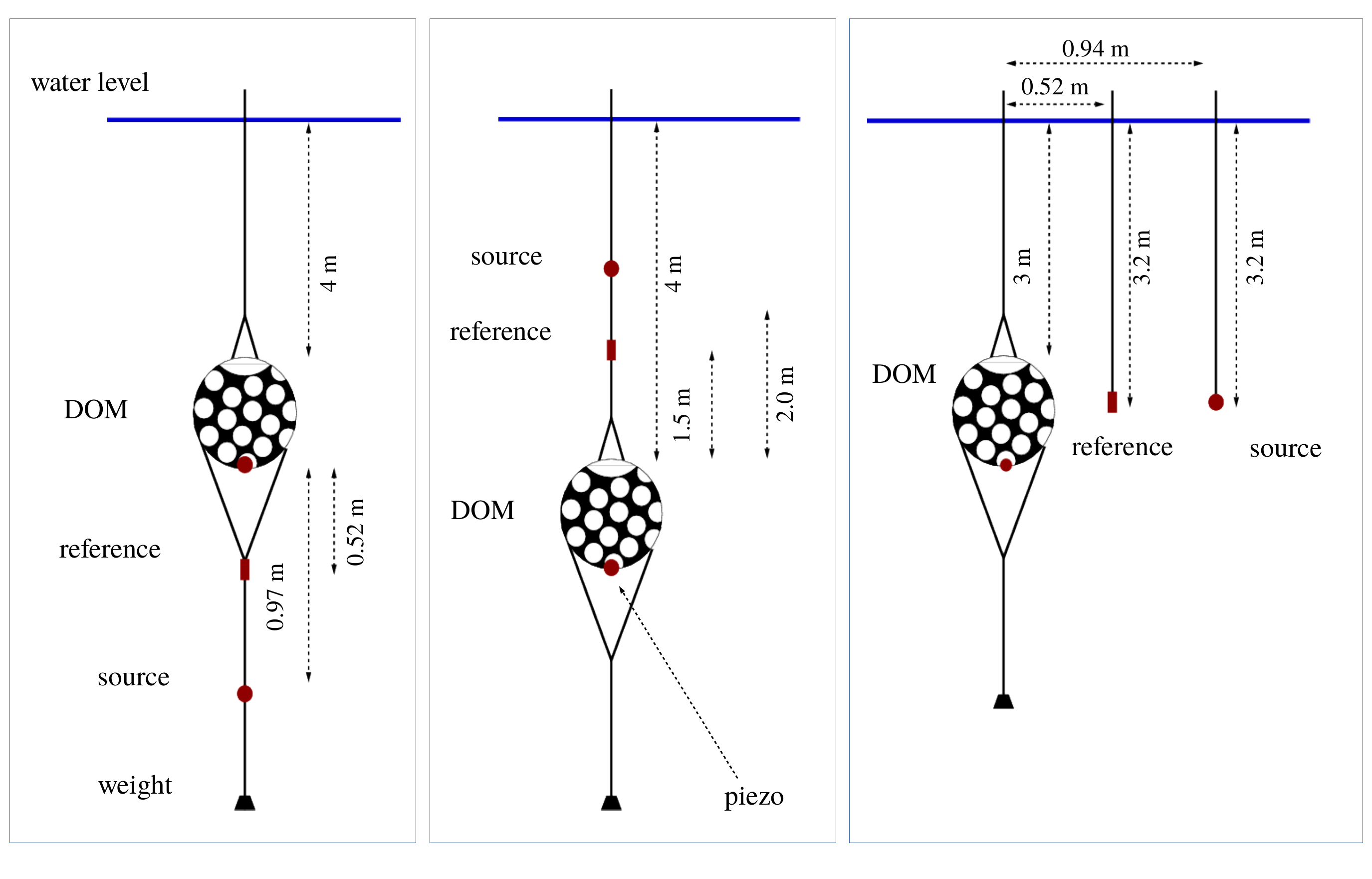}
      \caption{Three configurations of the relative position of the DOM, the reference hydrophone and the source. Note that in the middle panel the piezo hydrophone is indicated as shown in the picture in figure \ref{fig:KM3NeT_piezo}.}
      \label{fig:experimental_configs}
\end{figure}

In figure \ref{fig:experimental_configs} the experimental set up in the anechoic basin is shown. The DOM, reference hydrophone and the source were attached to a string and kept submerged by additional weight. The distances were chosen to minimize the influence of possible reflections from the water surface and from the bath bottom and walls. To compare the hydrophone under test data with the reference hydrophone data a distance correction was applied. Various configurations were used in which the DOM, the reference hydrophone and the source were placed at different relative positions. As the hydrophone is located on the 'south pole' of the DOM it is anticipated that the response depends on its orientation with respect to the source.
In figure \ref{fig:experimental_configs} three configurations are shown, with situations i) when the piezo directly faced the source, ii) when the piezo is on the backside of the DOM and iii) when the source is on the side of the DOM.

Next to the measurements in the anechoic basin, the front-electronics of the piezo hydrophone (separate from the DOM) was investigated to isolate non-acoustic effects. Results from the measurements carried out both in the basin as well as in the electronics laboratory are discussed in the next section.

\section{Measurement results}
\label{sec:results}

\subsection{Hydrophone front-end electronics characterization}
\label{sec:front-end}
\begin{figure}[h]
\centering
\begin{minipage}[]{0.45\textwidth}
\includegraphics[width=0.9\textwidth,clip]{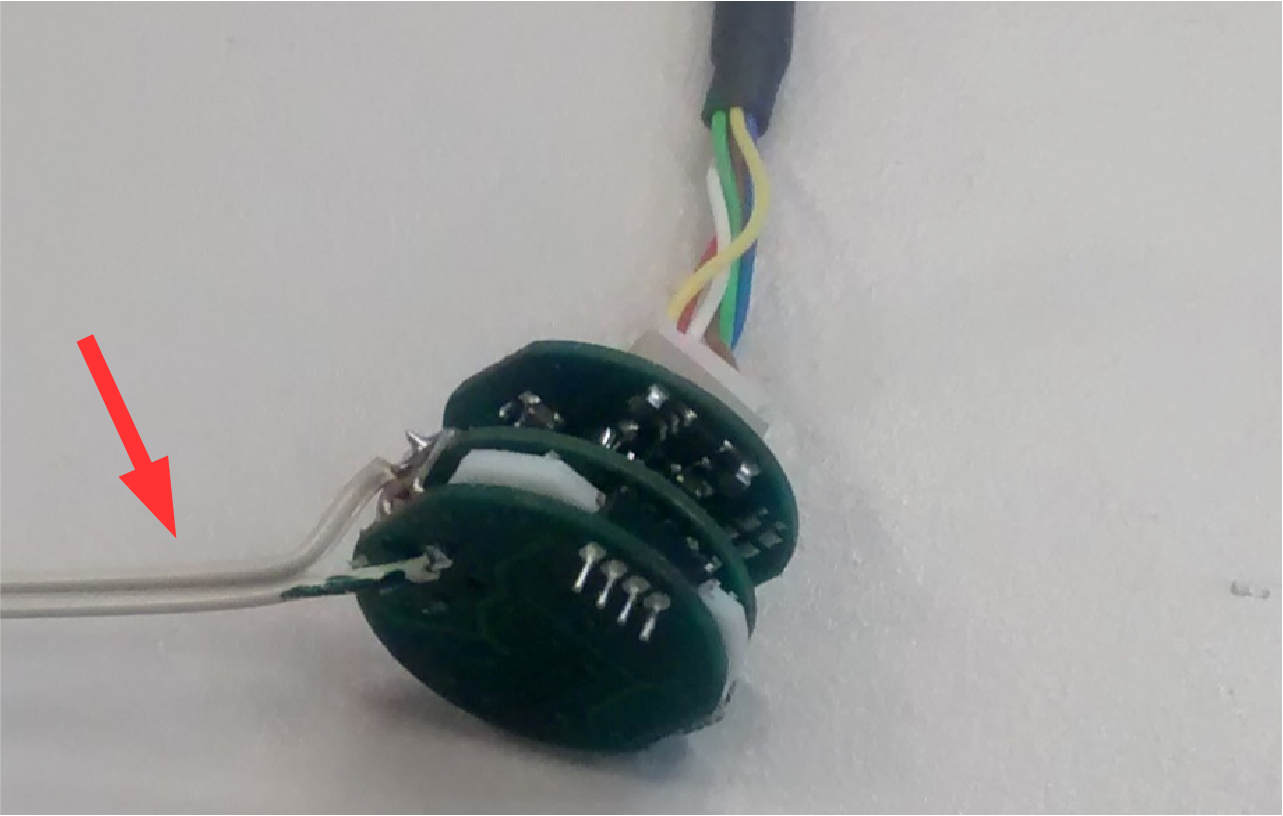}
\end{minipage}
\begin{minipage}[]{0.45\textwidth}
\includegraphics[width=0.95\textwidth,clip]{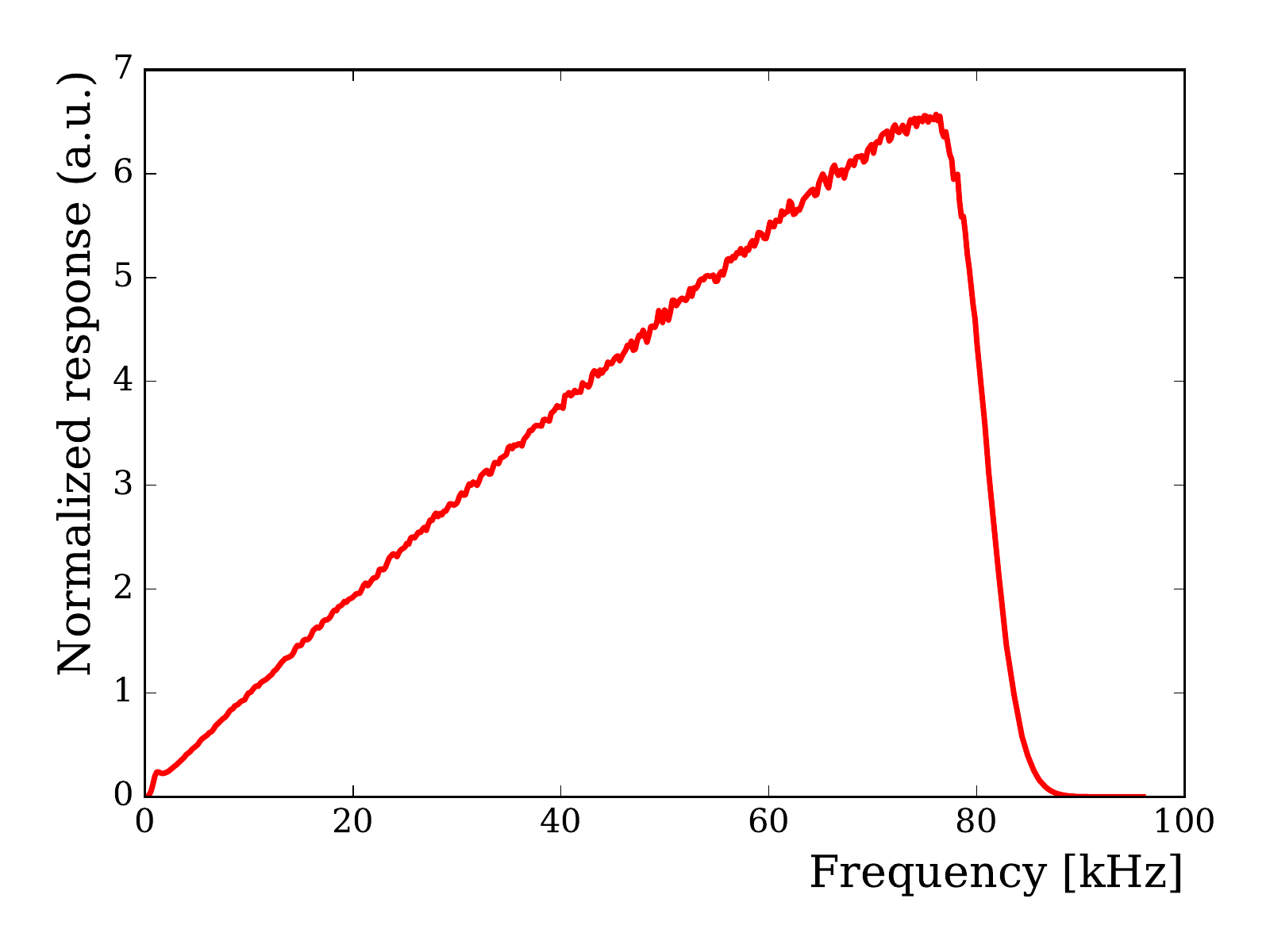}
\end{minipage}
\begin{minipage}[]{0.45\textwidth}
(a)
\end{minipage}
\begin{minipage}[]{0.45\textwidth}
(b)
\end{minipage}
\caption{(a) Photograph of the front-electronics shown with the piezo ceramic attached to it. The arrow indicates the external stimulus, i.e. a electrical connection to a signal generator In (b) the electronics response is shown as a function of the frequency. The response was normalized to the response at 10 kHz.}
\label{fig:electronics}
\end{figure}

To understand the (non-acoustic) response of the piezo hydrophone, the front-electronics have been subjected to characterization measurements without the piezo ceramic attached. A set-up was prepared in which the front-end electronics that is located on a small printed circuit board in close proximity of the piezo ceramic was stimulated using a signal from a function generator and read out using a laptop. A signal consisting of white noise between 1 and 80 kHz was offered to the electronics. The response was subsequently recorded and normalized to the response measured at 10 kHz. In figure \ref{fig:electronics} (a) a photograph of the printed circuit board that houses the front electronics is shown, while in (b) the measured response is shown. The response shows a clear linear increase, which has a slope of 0.054 kHz$^{-1}$.     

\subsection{Waveform characterization}
\label{sec:waveforms}
\begin{figure}[h]
  \centering
    \includegraphics[width=7 cm,clip]{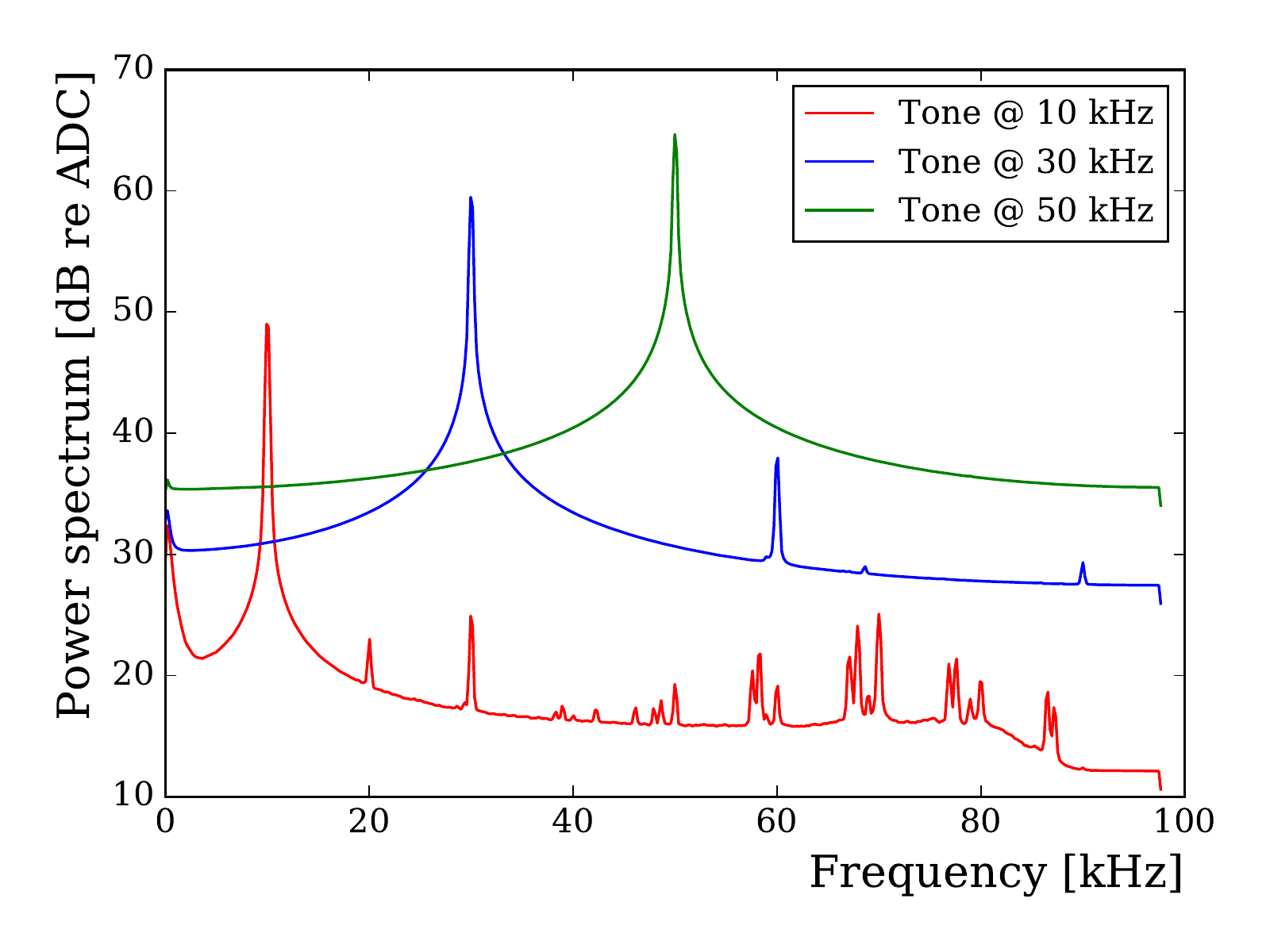}
  \caption{Examples of spectra of single tone signals at 10, 30 and 50 kHz as recorded using the DOM piezo hydrophone.}
    \label{fig:spectra}
\end{figure}

\begin{figure}[h]
  \centering
  \begin{minipage}[]{0.43\textwidth}
    \includegraphics[width=0.9\textwidth,clip]{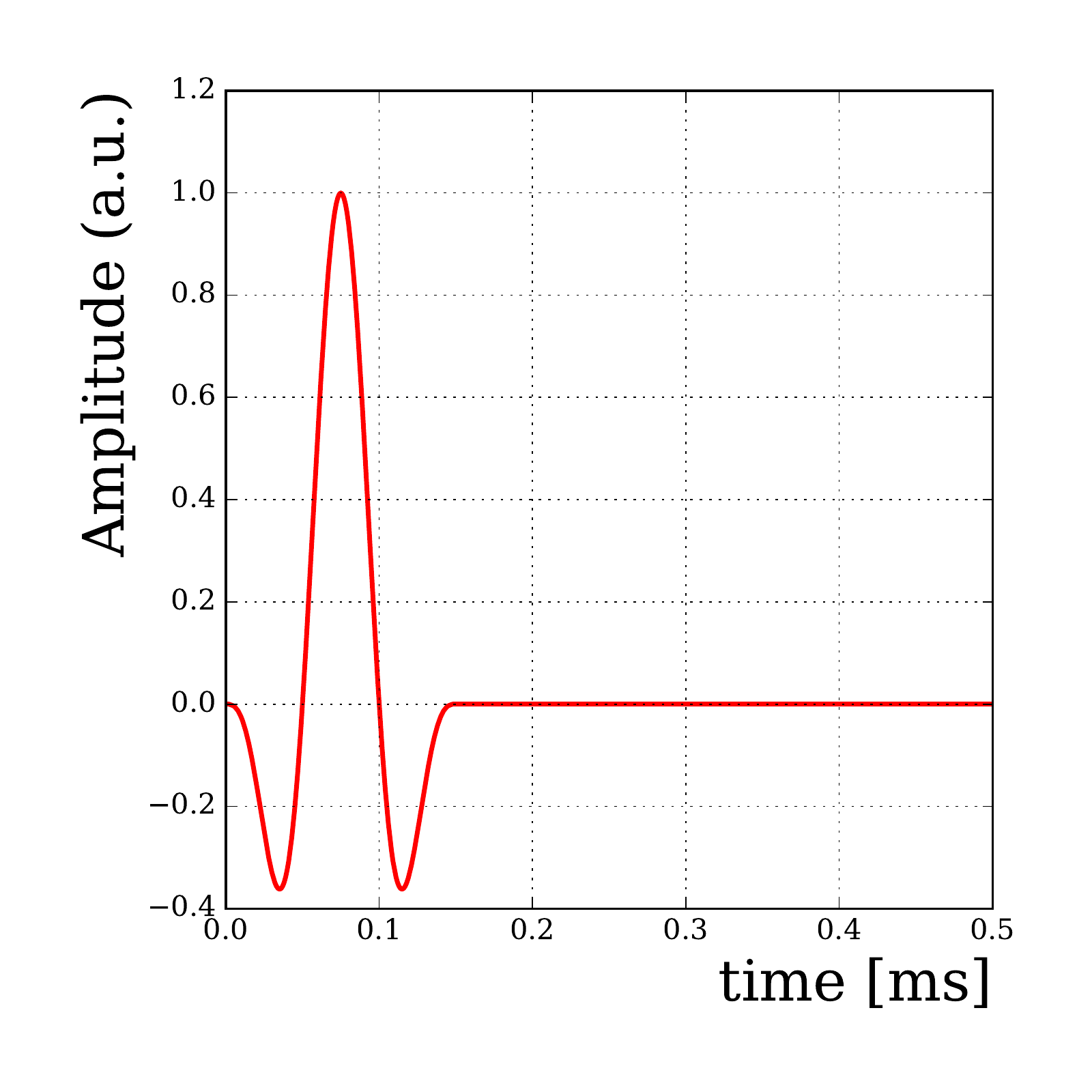}    
  \end{minipage}
  \begin{minipage}[]{0.56\textwidth}
    \includegraphics[width=0.9\textwidth,clip]{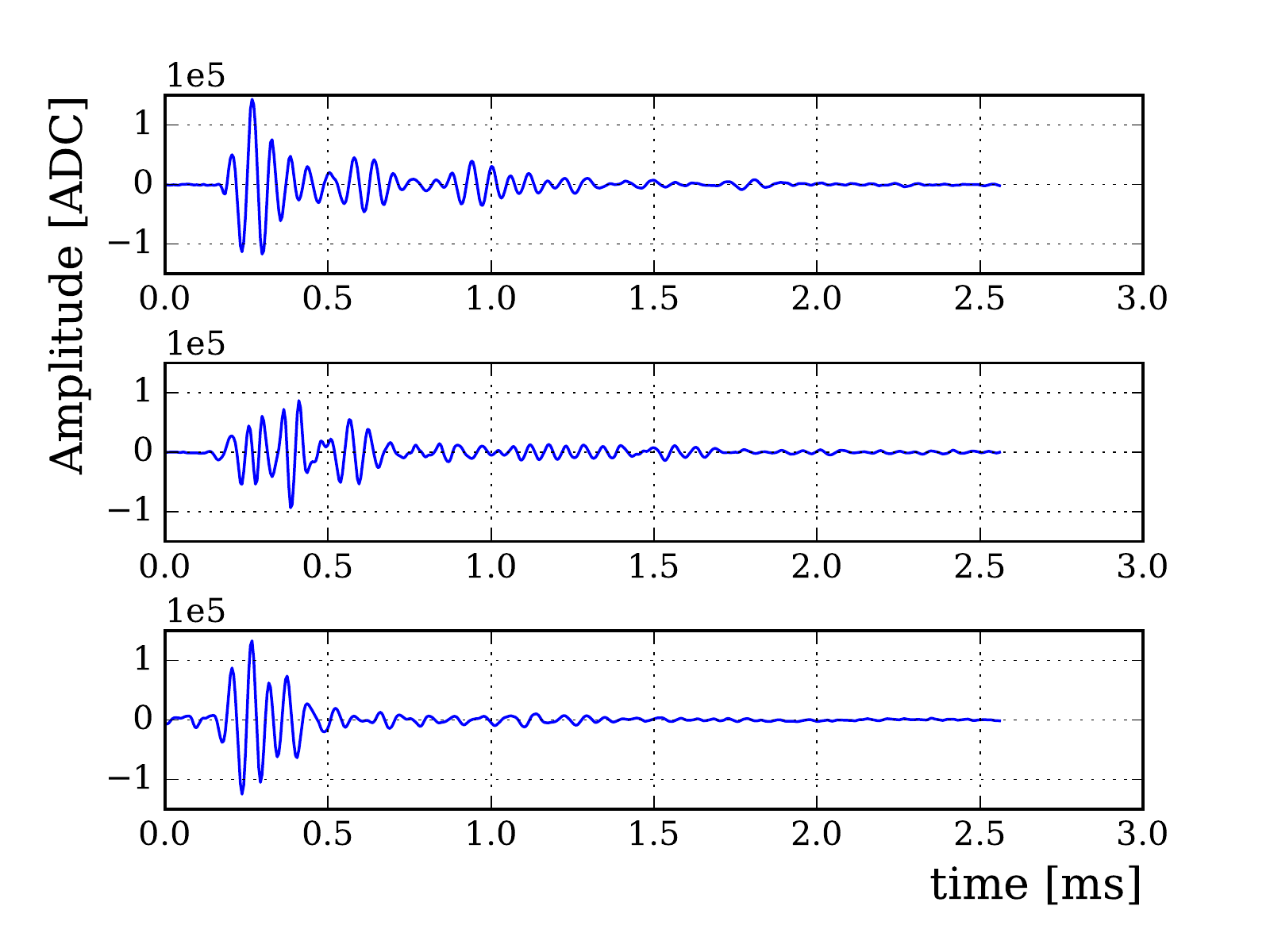}
  \end{minipage}
  \begin{minipage}[]{0.37\textwidth}
    \begin{center} (a) \end{center}
  \end{minipage}
  \begin{minipage}[]{0.56\textwidth}
    \begin{center} (b) \end{center}
  \end{minipage}
  \caption{(a) Examples of a signal as generated using a wave-generator. In (b) the corresponding waveforms are shown as recorded using the DOM hydrophone. The top, middle and bottom panels in (b) correspond to the experimental configuration in respectively the left, middle and right panel of figure \ref{fig:experimental_configs}.}
    \label{fig:wavelets}
\end{figure}

To investigate the acoustic response of the hydrophone in the DOM both continuous and transient waveforms were generated. Spectra of simple tones are shown in figure \ref{fig:spectra}, while in figure \ref{fig:wavelets} the time traces of wavelets, i.e. transients, are shown. In addition to the spectra and waveforms shown here, others were recorded such as frequency sweeps and bursts which are discussed in detail in \cite{Rasa18}. The spectra show a clear peak at the generated tone frequency and, in case of the 10 and 30 kHz tones, overtones are visible next to the fundamental peak. All power spectra in figure \ref{fig:spectra} show a typical 1/f noise at low frequencies, and the effect of the anti-alias filter at frequencies above 80 kHz \cite{Vio16}. Two effects cause a higher signal recorded at higher frequencies: As discussed in the previous section the electronic response is frequency dependent. There is an additional frequency dependence in the output of the source.

As an example of a generated transient waveform a wavelet is shown in figure \ref{fig:wavelets} (a). The recorded waveforms of 10 kHz wavelets are shown in \ref{fig:wavelets} (b) for three relative positions as explained in section \ref{sec:experimental}.
The response of the DOM piezo shows significant ringing (or echoing) in the time domain. This ringing is not noise, but is reproduced with every pulse in the time trace \cite{Rasa18}. This means that the DOM is acoustically a complex object that tends to deform a transient signal depending on its angle of arrival (as well as the main frequency component of the wavelet). It is important, however, to understand the ringing, as the angular dependence of the waveform will have impact on the performance of the acoustic positioning system (APS) that is used to determine the position of the DOM once deployed.

\subsection{Calibration and system noise}
\label{sec:calibration}
\begin{figure}[h]
  \centering
  \begin{minipage}[]{0.49\textwidth}
    \includegraphics[width=\textwidth, clip]{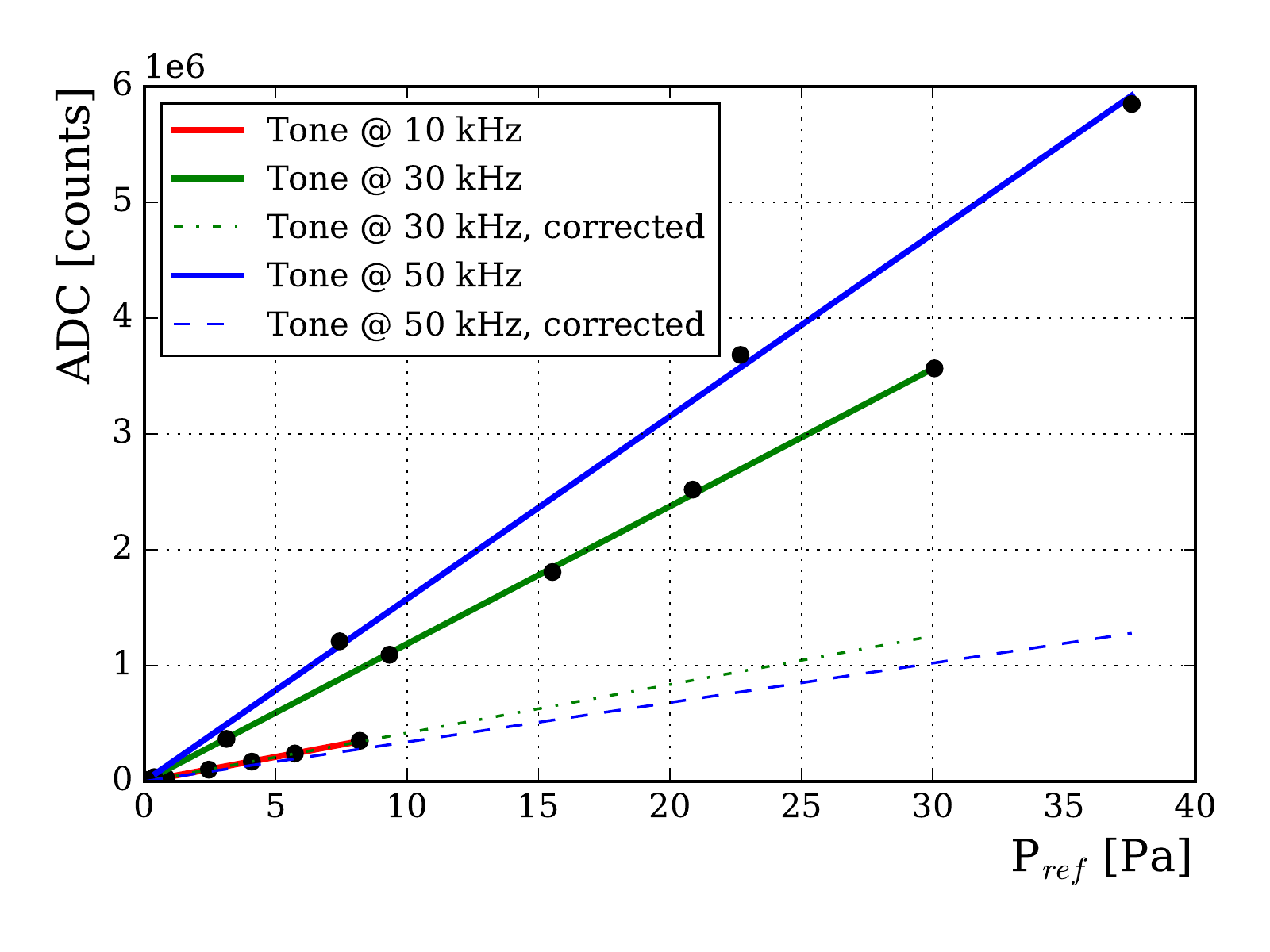}
  \end{minipage}
  \begin{minipage}[]{0.49\textwidth}
    \includegraphics[width=\textwidth, clip]{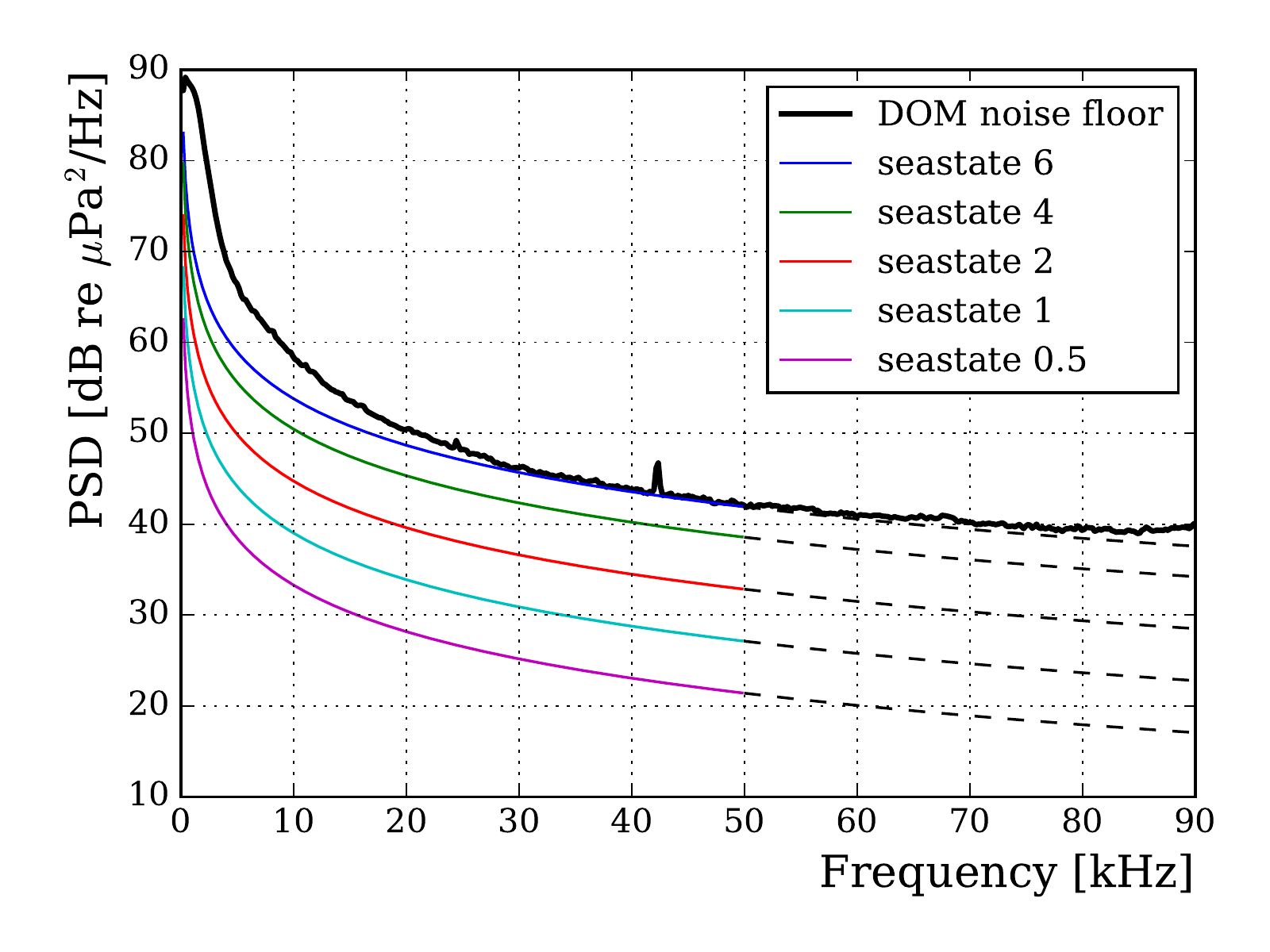}
  \end{minipage}
  \begin{minipage}[]{0.47\textwidth}
    \begin{center} (a) \end{center}
  \end{minipage}
  \begin{minipage}[]{0.47\textwidth}
    \begin{center} (b) \end{center}
  \end{minipage}
  \caption{(a) Calibration of the DOM piezo hydrophone for three tone frequencies. The dashed and dotted curves are the calibration curves corrected for the frequency dependent electronic response. The calibration procedure is explained in the text. In (b) the calibrated noise floor is shown, together with the definition of the sea state noise curves as defined in \cite{Hildebrand}.}
  \label{fig:calibration}
\end{figure}

To perform the signal height calibration, tones at various frequencies and at various sources levels have been recorded using both the DOM and reference hydrophone. For each recording the spectrum was integrated in a 1 kHz band around the peak of the tone and plotted in figure \ref{fig:calibration}. Straight lines have been fitted to the data to obtain calibration curves for three different tones, i.e. at 10, 30 and 50 kHz. As can be seen from figure \ref{fig:calibration}, the calibration curves show a linear behavior but have rather different slopes. This difference in slopes is explained by the frequency dependent response of the electronics. Values for the slopes are 42328, 118777, and 157610 ADC/Pa for the tones at 10, 30 and 50 kHz respectively. Note that when the calibration curves are corrected for the electronics response, the curves are overlapping as expected. The corrected calibration curves for the 30 and 50 kHz tones are overlaid in figure 6 as well (dashed and dotted lines) and show a reasonable agreement with the calibration curve for the 10 kHz tones.

Once the piezo hydrophone has been calibrated, the noise floor of the hydrophone can be determined. In figure \ref{fig:calibration}, the noise floor of the hydrophone is shown. 
The noise floor can be compared to the sea state noise as originally defined by Knudsen \cite{Knudsen} and reformulated in \cite{Hildebrand}. These sea state levels as drawn in figure \ref{fig:calibration} (b) are valid up to a frequency of about 50 kHz after which the thermal noise (not drawn) of the deep sea becomes the dominant source of noise. Above 30 kHz, the hydrophone noise corresponds to sea state 6.

\section{Conclusions}
\label{sec:conclusions}
In this paper we report on the characterization measurements carried out on the piezo hydrophone that are integrated in the Digital Optical Module (DOM) of the KM3NeT experiment. Measurements have been carried out on the front-end electronics in the electronics lab and on a complete DOM submerged in an anechoic basin. These measurements show that the hydrophone has an effective operational frequency range between 1 and 80 kHz. Continuous waveforms, such as tones, are well understood, but the response to transients show complex waveforms. There are a number of resonance frequencies, moreover the amplitude depends much on the position of the source with respect to the DOM. 
The hydrophone has been calibrated from which the minimum detectable acoustic pressure has been derived for a large range of frequencies.

Although the sensitivity of KM3NeT as an acoustic neutrino telescope might be limited, toothed whales and other yet unknown transients that might form a source of background in future dedicated acoustic neutrino telescopes, can be studied in great detail. To conduct such study, it is of large interest to implement a dedicated trigger for transient events in the data acquisition of KM3NeT. The transient response as presented in this paper could provide a basis for the definition of such a trigger. 

\section{Acknowledgements}
The authors would like to acknowledge M.~Ainslie for fruitful discussions and E.~Boer for his support during the measurements on the DOM.

\end{document}